\documentclass{revtex4}%
\usepackage{amsfonts}
\usepackage{amsmath}
\usepackage{amssymb}
\usepackage{graphicx}%
\begin{document}
\preprint{ }
\title[Short title for running header]{Discrimination and analysis of interactions in non-contact Scanning Force Microscopy}
\author{Elisa Palacios-Lid\'{o}n and Jaime Colchero}
\affiliation{Facultad de Quimica, Departamento de F\'{\i}sica, Universidad de Murcia,
E-30100 Murcia.}
\keywords{Scanning Force Microscopy, non contact tip-sample interaction, electrostatic
interaction, van der Waals interaction, nanometer scale characterisation and
analysis, force versus distance curves, 3D Mode, Force Volume.}
\pacs{07.50.-e; 07.79.Lh; 68.37.Ps; 73.40.Cg; 73.22-f;73.61-r}

\begin{abstract}
A method for the separation and quantitative characterization of the
electrostatic and Van der Waals contribution to tip-sample interaction in
non-contact Scanning Force Microscopy is presented. It is based on the
simultaneous measurement of cantilever deflection, oscillation amplitude and
frequency shift as a function of tip-sample voltage and tip-sample distance as
well as on advanced data processing. Data is acquired at a fixed lateral
position as interaction images\ with the bias voltage as fast scan and
tip-sample distance as slow scan. Due to the quadratic dependence of the
electrostatic interaction with tip-sample voltage the Van der Waals force can
be separated from the electrostatic force. Using appropriate data-processing
the Van der Waals interaction, the capacitance as well as the contact
potential can be determined as a function of tip-sample distance from the
force as well as from the frequency shift data. The measurement of resonance
frequency shift yields very high signal to noise ratio and the absolute
calibration of the measured quantities, while the acquisition of cantilever
deflection allows the determination of tip-sample distance. The separation and
quantitative analysis of Van der Waals and electrostatic interaction as
proposed in the present work results in precise and reproducible measurement
of tip-sample interaction that will significantly improve the interpretation
of SFM data and will substantially contribute to the characterization of
nanoscale properties by SFM.

\end{abstract}
\volumeyear{year}
\volumenumber{number}
\issuenumber{number}
\eid{identifier}
\date[Date text]{date}
\received[Received text]{date}

\revised[Revised text]{date}

\accepted[Accepted text]{date}

\published[Published text]{date}

\maketitle

\section{Introduction}

As scientific and technological interest focuses on increasingly smaller
length scales, tools for visualizing and characterizing nanoscale systems are
needed. An important technological step towards this goal is the development
of techniques that allow precise and quantitative analysis of materials on a
nanometer scale. In this context, Scanning Probe Microscopy, and in particular
Scanning Force Microscopy (SFM) \cite{SFMBinnig} has proved to be a very
powerful tool for Nanotechnology. SFM allows not only the visualization of
surfaces on a nanometer scale, but also its modification and the
characterization of material properties. SFM is based on the interaction of a
very sharp tip with the sample to be studied. Therefore, a deep understanding
of tip-sample interaction is fundamental in SFM. A variety of forces may act
between tip and sample: dispersion, electrostatic, chemical, elastic as well
as adhesion and friction forces. In addition, when tip and sample are composed
of magnetic materials, also magnetic forces act. Each of these forces in
principle opens the way for measuring the corresponding physical property of
the sample.

The variety of different forces - and thus different material properties -
that can be measured is one of the reasons for the great versatility of SFM.
Correspondingly, at present SFM is used very successfully and extensively in a
variety of scientific areas. However, we believe that it is still not the
precise and quantitative tool required by the Nanotechnology community. One of
the difficulties for quantitative measurements with SFM is -in our opinion-
precisely the wealth of interactions that may act in a SFM set up. Since, a
priory, only the total force is measured, it is difficult to discriminate
between the contributions of the different kind of forces. However, to obtain
quantitative measurements and for a complete characterization of material
properties on an nanometer scale, as well as in general for the correct
interpretation of SFM experiments, the determination of the origin and the
relative strength of the measured forces is fundamental.

Due to its importance, tip-sample interaction has been the topic of a variety
of studies. Some of these studies are focused on the modeling and analysis of
the dynamics of the oscillating tip within the (non-linear) surface
potential\cite{TappingAncowski,TappingOthmar,TappingGissibl,TappingAlvaro},
while others are devoted to the physics of the interaction
itself\cite{InteractionChen,Argento,InteractionStark}. In this context, the
modeling and measurement of electrostatic interaction within a SFM set-up has
received particular
attention\cite{ESFMTipConeMole,ESFMTipCone,moleAPLnew,ESFMReview}. A variety
of methods have been used to experimentally characterize the interaction
between tip and sample in a SFM-setup. The simplest is based on the
acquisition of force versus distance (\textit{FvsD})
curves\cite{FvsDWeissenhorn}. This mode measures the static deflection of the
cantilever as tip-sample distance is varied and is easy to implement but is
not sensitive enough to analyze weak forces. A more precise method measures
variations of the oscillation of the cantilever, that is, of its amplitude,
phase or resonance frequency (amplitude, phase or frequency vs. distance
curves). Dynamic methods are more sensitive due to signal enhancement induced
by the resonance of the cantilever\cite{ResonanceAlbrecht}. More sophisticated
methods are based on the multidimensional acquisition of data, where the
interaction is measured not only as a function of tip-sample distance, but as
a function of at least a second variable. In one of those modes the mechanical
spectrum of the cantilever is determined as a function of tip-sample distance,
in this case the second variable is the excitation frequency. The oscillation
of the tip is induced either by thermal fluctuation\cite{DuerigResNoise}, or
by external excitation\cite{InteractionDucker,InteractionRon,InteracionYo}.

SFM experiments can be performed in Ultra High Vacuum (UHV) conditions or in
liquids and air. While the first kind of experiments are more demanding, the
latter are more difficult to understand due to less controlled surface
conditions: in air adsorbed liquid films and different kind of contamination
may complicate data interpretation. Moreover, while in UHV conditions the tip
can approach the surface as near as atomic distances and "feel" chemical
interactions, in air the minimum tip-sample distance is about one magnitude
larger, since the tip snaps to the surface not due to the instability produced
by typical surface potentials, but due to the (spontaneous) formation of
liquid necks\cite{InteracionYo}, occurring at distances between 2 and 5 nm.
Correspondingly, quantitative determination of tip-sample interaction in air
is more challenging since signals are typically weaker and interaction data
has to be analyzed and interpreted with
care\cite{CuellosMonica,CuellosHerminghaus}. In many cases the correct
determination of tip-sample distance and of the "true" non-contact regime is
crucial\cite{CommentNonContact1}.

In the "true" non-contact regime dispersion forces, which are always present,
and electrostatic forces, which are the strongest forces and have long
interaction range, are the most relevant forces in a typical SFM set up.
Separation of these two kind of interactions is fundamental for a better
understanding of tip-sample interaction, for adjusting of optimum imaging
conditions as well as for the quantitative determination of material
properties on a nanometer scale. In principle, as recognized in previous
works, variation of tip-sample bias results in turning \textquotedblleft
on\textquotedblright\ and \textquotedblleft off\textquotedblright\ the
electrostatic interaction, allowing to separate the dispersion interaction
from the electrostatic interaction. In fact, to experimentally verify the
quadratic dependence of tip-sample interaction on on tip-sample voltage and to
characterize the dielectric properties of different samples Hu et
al.\cite{MiguelAPL} have acquired interaction versus voltage curves. In a
similar approach, Guggisberg et al.\cite{MeyerComparisonInteractions} have
measured tip-sample interaction as a function of tip-sample voltage and
tip-sample distance to determine and compensate the contact potential between
tip and sample. The goal of that work performed in ultra high vaccum was the
precise discrimination and control of dispersion and electrostatic
interactions in order to measure short ranged chemical forces. In the present
work we pretend to further develop these techniques for separation and
measurement of electrostatic and dispersion interaction. As in the works just
discussed data is acquired as a function of tip-sample distance and of
tip-sample voltage. The corresponding experimental data sets are stored and
visualized as "interaction images". The to obtain a complete characterization
of tip-sample interaction cantilever deflection (force), oscillation amplitude
as well as frequency shift are acquired simultaneously and processed using
appropriate data processing algorithms. The measurement of resonance frequency
shift yields very high signal to noise ratio and the absolute calibration of
the measured quantities, acquisition of oscillation amplitude allows to
recognize the "true" non-contact regime, and from the cantilever deflection
tip-sample distance is determined. Precise values for the dispersion
interaction, the contact potential and the tip-sample
capacity\cite{CommentCapacity} are obtained as a function of tip-sample
distance. Furthermore, this method allows to characterize parameters such as
the tip radius and the tip-sample distance. We note that this method requires
no previous information about the electric properties of the tip-sample system
-and in particular about the contact potential- since this information is
obtained \textquotedblleft self-consistently\textquotedblright\ by the
algorithm processing the interaction images. In particular for experiments
performed in air, but also for UHV applications we are convinced that the
method presented in this work will result in improved data acquisition and
data interpretation. Moreover, we believe that this method will contribute to
obtain the profound understanding of tip-sample interaction needed to
quantitatively determine electrostatic properties of samples on a true
nanometer scale.

\section{Theoretical Background}

In a typical SFM experiment tip-sample interaction $I(d)$ (that is, energy) is
not measured directly. Instead the force $F(d)=-I^{\prime}(d)$ or the
resonance frequency are determined from experiment. The resonant frequency of
the tip-sample system and the curvature $I^{\prime\prime}(d)$ of the
interaction potential are related by%
\begin{equation}
\upsilon_{0}(d)=\sqrt{\frac{c_{lev}+I^{\prime\prime}(d)}{m_{eff}}}%
/(2\pi)\simeq\upsilon_{00}(1+\frac{1}{2}\frac{I^{\prime\prime}(d)}{c_{lev}})
\label{ResFreq}%
\end{equation}
with $\upsilon_{00}$ free resonance frequency of the cantilever, $c_{lev}$ its
force constant and $m_{eff}$ the effective mass of the cantilever. We note
that this relation is only correct if the oscillation amplitude is
sufficiently small in order to avoid non-linearities of the
interaction\cite{FrequencyApprox}. The approximation is valid as long as
$I^{\prime\prime}(d)\ll c_{lev}$ or, equivalently, for small shifts
$\Delta\upsilon_{0}(d)=\upsilon_{00}-\upsilon_{0}(d)\ll\upsilon_{00}$ of the
resonance frequency. For a detailed discussion of the mechanical behavior of a
SFM setup see Duerig\cite{InteractionDuerig}.

To simplify the argumentation, in the context of the present work we will
assume a homogeneous sample. We note, however, that the results discussed here
are also relevant for heterogeneous samples as long as the sample is "locally"
homogeneous, which means that\ the material properties of the sample are
homogeneous on a length scale larger than the resolution of the SFM-system,
that is, its effective aperture function. We will also assume that in the true
non-contact regime\cite{CommentNonContact1} tip-sample interaction is governed
only by dispersion and electrostatic interaction:%

\begin{equation}
I(d)=I^{dis}(d)+I^{estat}(d) \label{eqInteraction}%
\end{equation}
with $I^{estat}(d)$ electrostatic interaction and $I^{dis}(d)$\ dispersion
interaction. From a fundamental point of view, both kind of interactions can
be considered ultimately of electronic origin. While the electrostatic
interaction arises "directly" from charges, dispersion forces arise
"indirectly" through the residual interaction of fluctuating dipoles within
matter (Van der Waals forces) or fluctuating electric fields in vacuum
(Casimir forces)\cite{DispersionGeneral}. Van der Waals and Casimir forces are
of quantum mechanical origin. In the context of\ the present work, we will
assume that the dispersion forces within a SFM-setup are well described by the
relation\cite{Argento,Israelachvili}%
\begin{equation}
F^{vdW}(d)=\frac{A_{tms}~R}{6~d^{2}} \label{VanDerWaals}%
\end{equation}
which describes the Van der Waals force between a tip of radius R and the
sample surface by means of the Hamaker constant $A_{tms}$\ ($t$ip interacting
with $s$ample through $m$edium, see Israelachvili\cite{Israelachvili} for a
detailed discussion).

If tip and sample are electrical conductors the electrostatic force between
tip and sample can be written as a surface integral over the electric field on
the sample surface\cite{Jackson}. In addition, the electric field lines can be
approximated by segments of circles connecting tip and sample. To a good
approximation, the electric potential decays linearly along these circular
segments. Within this approach, the electrostatic force between tip and sample
can be calculated as \cite{Guthman}:
\begin{equation}
F\left(  d\right)  =\int_{S}dS\,\frac{\varepsilon_{0}}{2}\;E\left(
x,y,d\right)  ^{2}\simeq\frac{\varepsilon_{0}U_{0}^{2}}{2}\int_{S}dS\,\frac
{1}{a(x,y,d)^{2}} \label{eqForce}%
\end{equation}
where $E(x,y,d)$ is the electric field on the surface for a certain tip-sample
distance $d$, $U_{0}$ is the effective voltage between tip and sample and
$a(x,y,d)$ the arc length of the circular segment coming from the probe and
ending on a point $(x,y)$ of the surface. When nanoscale dielectric systems
are adsorbed on a conducting surface, it can be shown within a perturbative
approach\cite{moleperturbative} that tip-sample interaction depends also on
the electrical properties of these systems, which are the objects to be
characterized by SFM.

Many SFM experiments are performed with a conducting or semiconducting tip in
order to control its potential. It would seem reasonable to assume that
grounding the tip with respect to the sample surface would imply vanishing of
electrostatic forces. This is, however, only correct if the work function (or
contact potential) of tip and sample are equal. Otherwise, differences in work
function induce transfer of charges that result in electrostatic fields and
thus, according to equation \ref{eqForce}, in electrical forces even if no
external bias is applied. The electrostatic force between tip and sample is
therefore a quadratic function of tip voltage with its minimum shifted by an
amount $U_{cp}$ with respect to the origin due to contact potential
difference\cite{WickramasingeKPM,KPMAbraham}:%
\begin{equation}
F_{el}(d)=C^{\prime}(d)(U_{tip}-U_{cp})^{2}/2 \label{SFMforce}%
\end{equation}
where $C^{\prime}(d)$ is the derivative of the tip-sample capacity. For a
conductive tip and sample, this derivative of the\ capacitance can be
estimated by the surface integral of equation \ref{eqForce}%
\begin{equation}
C^{\prime}(d)=\varepsilon_{0}\int_{S}dS\,\frac{1}{a(x,y,d)^{2}}
\label{approxCapacitance}%
\end{equation}

For the case of Van der Waals interaction we assume a specific tip geometry
-namely a parabola described by a tip radius $R$ , see relation
\ref{VanDerWaals}- while for the case of the electrostatic interaction
equation \ref{eqForce} is generic, that is, in principle any tip-sample
geometry can be described using an appropriate tip-sample capacity $C(d)$.
This different treatment is justified because of the different distance
dependencies of Van der Waals and electrostatic interactions. Due to its
faster decay Van der Waals interaction is less sensitive to the large scale
geometry of the probe (see Argento et. al.\cite{Argento} and\ Colchero et.
al.\cite{modeloPRB} for a more detailed discussion). In fact, for the
experiments described in the present work we find that the commonly used
approximation - $C(d)=2\pi~\varepsilon_{0}~R\ln(d)$ - for the tip-sample
capacitance\cite{ESFMSpere} does not describe tip-sample interaction
satisfactory within our experimental error. Therefore, the relation%
\begin{equation}
C(d)=2\pi\varepsilon_{0}R\ln\left(  \frac{d}{d+R(1-\sin\vartheta_{0})}\right)
\label{CapacityComplex}%
\end{equation}
for the tip-sample capacity will be used, which results from modeling the tip
as a truncated cone of opening angle $\vartheta_{0}$ ending smoothly in a
spherical tip apex of radius $R$ (see Hudlet et. al. \cite{Guthman} for a
detailed discussion of this model probe tip).

As a general result from this theoretical section we conclude that for a fixed
distance $d_{0}$ the total tip-sample interaction as a function of voltage and
tip-sample distance is of the form%
\begin{equation}
i(U,d_{0})=\alpha(d_{0})+\gamma(d_{0})(U-\kappa(d_{0}))^{2}/2
\label{eqParabola}%
\end{equation}
where the constant term $\alpha$ describes the Van der Waals interaction,
$\gamma$\ the curvature of the parabola induced by electrostatic interaction
and $\kappa$ the position of the minimum. Note that this general relation
applies to the force as well as for the force gradient, and thus also for the
resonance frequency. This quadratic dependence of tip-sample interaction on
voltage is essentially the basis for the separation of Van der Waals and
electrostatic interaction in this work (see also Hu et al. \cite{MiguelAPL} as
well as Guggisberg et al. \cite{MeyerComparisonInteractions}).

\section{Experimental Method}

Usually either the force, the resonance frequency or the oscillation amplitude
are measured to characterize the tip-sample interaction. In the present work,
all three quantities are measured simultaneously. By measuring the force the
first derivative of the interaction (equation \ref{eqInteraction}) is
determined experimentally while the resonance frequency is related to the
second derivative of the interaction. As will be discussed in more detail
below, the method presented here allows the precise determination of all
parameters relevant for tip-sample interaction. By measuring a family of
curves $i(U,d_{0})$ as a function not only of tip voltage $U$ but also of
tip-sample distance \textit{interaction images}\ $i(U,d)$ can be obtained.
Each horizontal line can then be adjusted to the parabola defined by equation
\ref{eqParabola} to obtain, for each tip-sample distance, the contribution of
the Van der Waals interaction, the contact potential and the tip-sample
capacitance. From a whole \textit{interaction image}\ $i(U,d)$\ three curves
are obtained, one, $\alpha(d)$, characterizing the Van der Waals interaction
as a function of distance, another one, $\gamma(d)$, describing the
capacitance of the tip-sample system as a function of distance and the third
one, $\kappa(d)$, corresponding to the contact potential as a function of
tip-sample distance.

\subsection{Data acquisition}

Figure 1 schematically shows the experimental implementation of the setup used
to acquire the \textit{interaction images}. The experiments were performed
with a NanoTec SFM system composed of SFM head, high voltage controller and
PLL/dynamic measurement board\cite{nanotec}. For our experiments, we find that
cantilevers with long tips, located at its very end are ideal, since these
kind of SFM\ probes minimize the effect of the interaction between the sample
and the (macroscopic) cantilever, which can result in uncontrolled
electrostatic interaction\cite{EstatNanotechnology} and (long-range) viscous
forces\cite{LangmuirViscoso} that decrease the quality factor of the
oscillation and thus the sensitivity in dynamic detection modes.\ Olympus
OMCL-AC-type cantilevers have been used in all experiments\cite{Cantilever}.
The \textit{interaction images}\ were acquired using the
"3D-Mode"\cite{3Dmodes}, which allows fast switching from normal imaging to a
series of extended acquisition modes. "3D-Mode" is a generalized acquisition
mode where data -that is, some input channel- is acquired as a function of two
output channels that are user selected: $datain=datain(output_{1},output_{2}%
)$. "3D-Mode" images are acquired in a raster scan mode where one channel is
varied "fast" at a fixed value of the "slow" output, then the "slow" output is
varied by one step and a new line of data is acquired by varying the "fast"
output. For the experimental data shown here, the \textquotedblleft%
3D-Mode\textquotedblright\ was set in order to have the tip voltage as fast
scan and the tip-sample distance as slow scan, that is, the tip-sample voltage
is ramped fast to acquire, for each distance, interaction data as a function
of tip-sample voltage while tip-sample distance is varied slowly to bring the
tip into and out of contact with the sample\cite{PNAS,Estat3D}. For each set
of experiments, the normal force, the frequency shift and the oscillation
amplitude are measured as a function of tip-sample voltage and tip-sample distance.

To measure the resonance frequency of the tip-sample system the cantilever is
excited near its resonance frequency by a small piezoelectric element and a
Phase Locked Loop (PLL)\cite{ResonanceAlbrecht} is used to track the resonance
frequency as bias voltage and tip-sample distance are varied. The use of a
PLL-circuit allows the direct measurement of the resonance frequency in Hertz,
and thus the determination of interaction data in physically meaningful units.
Calibration of force data was performed using the nominal force constant of
the cantilevers to convert deflection into force and by calibrating the
photodiode signal with a \textit{FvsD} curve\cite{CommentCali}. Typical
oscillation amplitudes were 0.1 to 2.5$%
%TCIMACRO{\unit{nm}}%
%BeginExpansion
\operatorname{nm}%
%EndExpansion
$ (0.2 to 5$%
%TCIMACRO{\unit{nm}}%
%BeginExpansion
\operatorname{nm}%
%EndExpansion
$ peak to peak). The lower threshold of the oscillation is set by thermal
noise of the cantilever, while the higher value was chosen in order to avoid
non-linearities of the interaction. As a reasonable criterion we have chosen
amplitudes that are smaller than the minimum tip-sample distance at the snap
to contact point.

\subsection{Data processing}

Data is processed as follows: in a first step, each force line is averaged to
obtain a \textit{FvsD} curve. This \textit{FvsD} curve is then analyzed by
means of an appropriate algorithm that determines the snap to contact point in
order to separate the contact and non-contact regimes of the curve. We note
that, since data is acquired in air, the snapping instability is induced by
the spontaneous condensation of a liquid neck between tip and sample at rather
large distances of 2-5$%
%TCIMACRO{\unit{nm}}%
%BeginExpansion
\operatorname{nm}%
%EndExpansion
$\cite{InteracionYo,CuellosMonica}. In order to obtain meaningful interaction
data the jump distance, which can only be measured if the cantilever
deflection (force) is acquired, has to be taken into account.\ From the
\textit{FvsD} curve the snapping distance is measured and the position of the
surface as well as the true tip-sample distance is determined. All curves
calculated from the same set of \textit{interaction images}\ are then shifted
horizontally by a constant amount corresponding to the position of the surface
as "seen" by the \textit{FvsD} curve. Then, the origin of the horizontal axis
(piezo displacement) corresponds to the displacement of the piezo where the
cantilever would touch the surface if no attractive forces where present. In a
second step, force and frequency data corresponding to the non-contact regime,
that is, before the snapping instability, are adjusted to a quadratic function
according to equation \ref{eqParabola}. From the force \textit{interaction
image} three curves $f^{vdW}(d)$, $C^{\prime}(d)$ and $U_{cp}^{force}(d)$ are
obtained describing the Van der Waals force, the first derivative of the
tip-sample capacitance, and the contact potential as "seen" by the force data.
In the same way the frequency \textit{interaction image} yields three
corresponding curves $\Delta\omega_{0}^{vdW}(d)$, $C^{\prime\prime}(d)$ and
$U_{cp}^{freq}(d)$; the frequency shift induced by the Van der Waals
interaction, the second derivative of the capacitance and the contact
potential as "seen" by the frequency data. Finally, the three most relevant
curves - $\Delta\omega_{0}^{vdW}(d)$, $C^{\prime}(d)$ and $C^{\prime\prime
}(d)$ - are compared with the relations%
\begin{equation}
\Delta\omega_{0}^{vdW}(d)=\frac{1}{2}\frac{\omega_{00}}{c_{lev}}%
V_{vdW}^{\prime\prime}(d)=\frac{1}{6}\frac{\omega_{00}}{c_{lev}}%
\frac{A~R_{vdW}}{(d-d_{0}^{vdW})^{3}} \label{freqVdW}%
\end{equation}

\begin{equation}
C^{\prime}(d)=2\pi\varepsilon_{0}R_{estat}\frac{R_{estat}(1-\sin\left(
\vartheta\right)  )}{(d-d_{0}^{est1})\left(  d-d_{0}^{est1}+R_{estat}%
(1-\sin\vartheta\right)  }+b_{1} \label{CapacityForce}%
\end{equation}%
\begin{equation}
C^{\prime\prime}(d)=2\pi\varepsilon_{0}R_{estat}\left(  \frac{1}%
{(d-d_{0}^{est2})^{2}}-\frac{1}{\left(  d-d_{0}^{est2}+R_{estat}%
(1-\sin\vartheta)\right)  ^{2}}\right)  +b_{2} \label{CapacityFrequency}%
\end{equation}
which describe the frequency shift expected for a pure Van der Waals
interaction of a tip with radius $R_{vdW}$, and the first and second
derivatives of the capacitance predicted for a truncated cone ending in a
spherical tip apex of radius $R_{estat}$. The two constants $b_{1}$ and
$b_{2}$ are introduced to account for the possible effect of long range
contributions to the electrostatic interaction due to the tip cone or the
cantilever (see \cite{EstatNanotechnology,EstatSadewasser,MoleAPLTipConstant}%
). Each adjustment is allowed to find its own tip radius. In addition, each
adjustment is allowed to find the best value for $d_{0}$. Note that the
parameters $d_{0}$ represent mathematically the poles of the interaction, that
is, the position where the interaction diverges. From a physical point of
view, these poles correspond to the position of the surface. Complete
agreement with the surface position obtained from the \textit{FvsD} curve
would require $d_{0}=0$, since, as described previously, all curves have been
shifted to this position. From a physical point of view different tip radii
$R$ and different distances $d_{0}$ imply that the corresponding curves
\textquotedblleft see\textquotedblright\ different tips and different surface positions.

\section{Experimental Results}

Measurements have been performed on a variety of samples under different
experimental conditions. As representative cases two experiments performed on
a Au(111) surface are presented here, one taken with a highly doped silicon
tip at a relatively large oscillation amplitude of 2.5$%
%TCIMACRO{\unit{nm}}%
%BeginExpansion
\operatorname{nm}%
%EndExpansion
$ (5$%
%TCIMACRO{\unit{nm}}%
%BeginExpansion
\operatorname{nm}%
%EndExpansion
$ peak to peak), and another one taken with a Pt-coated tip at a significantly
lower amplitude of 1$%
%TCIMACRO{\unit{nm}}%
%BeginExpansion
\operatorname{nm}%
%EndExpansion
$. We consider the first case to be representative for typical experiments in
SFM and Electrostatic Force Microscopy, while the second case is closer to the
ideal situation where tip and sample are truly metal surfaces and the
oscillation amplitude is sufficiently small to neglect non-linear effects in
the tip-sample interaction.

Figure 2 shows a typical set of force and frequency \textit{interaction
images} as well as an oscillation amplitude image acquired on an Au(111)
surface with a Si tip\cite{Cantilever} in ambient air (temperature $\simeq$22$%
%TCIMACRO{\U{b0}}%
%BeginExpansion
{{}^\circ}%
%EndExpansion
$C, relative humidity $\simeq$50\%). These images show pure raw data, no plane
or other form of data-processing has been applied. The lower part of the
images corresponds to small tip-sample distances (near the surface), the
lowest lines that appear flat correspond to data taken with tip and sample in
mechanical contact. In this regime, none of the \textit{interaction images}
show the quadratic dependence on the tip-sample voltage that is visible in the
non-contact regime. The frequency shift image is completely saturated because
the elastic interaction drives the resonance frequency of the tip-sample
system far away from the free resonance frequency\cite{CommentLockRange}. In
the normal force the\ quadratic dependence is observed for small and large
tip-sample distances, while in the\ amplitude and in the frequency
\textit{interaction image}\ the quadratic dependence is observed only very
near the surface ($\lesssim$10$%
%TCIMACRO{\unit{nm}}%
%BeginExpansion
\operatorname{nm}%
%EndExpansion
$). As will be discussed in more detail below, this indicates a very short
interaction range for the frequency and the amplitude signal. The parabolic
dependence of the amplitude is attributed to increased dissipation in the
tip-sample system, but at the moment the precise origin of this dissipation is
still unknown. In the normal force image the jump to contact is recognized as
a step towards lower force values followed by a continues increase of the
normal force.

Figure 3 shows the \textit{FvsD} curve obtained by averaging each line of the
force \textit{interaction image}. In addition, the amplitude image has been
processed to calculate the oscillation amplitude for each tip-sample distance,
which allows to determine the upper and lower turning point of the tip motion
during oscillation shown in figure 3. As discussed above, the oscillation
amplitude ($\simeq2.5%
%TCIMACRO{\unit{nm}}%
%BeginExpansion
\operatorname{nm}%
%EndExpansion
$) has been chosen to be smaller than the distance corresponding to the jump
to contact regime. The mean jump distance measured from the force curve is
about $\overline{d}_{jump}\simeq4%
%TCIMACRO{\unit{nm}}%
%BeginExpansion
\operatorname{nm}%
%EndExpansion
$, however, we believe that the effective jump distance is
\[
d_{eff}=\overline{d}_{jump}-a_{osci}%
\]
where $a_{osci}\simeq2%
%TCIMACRO{\unit{nm}}%
%BeginExpansion
\operatorname{nm}%
%EndExpansion
$ is the oscillation amplitude just before the jump to contact instability.
The effective jump distance is then about 2$%
%TCIMACRO{\unit{nm}}%
%BeginExpansion
\operatorname{nm}%
%EndExpansion
$. The reasoning behind this correction is that the jump to contact is
believed to be induced by condensation of liqid necks at the lowest turning
point of the oscillation, that is, at the smallest tip-sample
distance\cite{InteracionYo}. Jump to contact then occurs at the smallest
tip-sample distance rather than from the mean tip-sample distance. The force
curve shows very low adhesion and very little hysteresis: the area enclosed
during the whole forward and backward cycle is the energy dissipated during
the acquisition process and corresponds to about $10^{-17}$~$%
%TCIMACRO{\unit{J}}%
%BeginExpansion
\operatorname{J}%
%EndExpansion
$ ($\simeq60%
%TCIMACRO{\unit{eV}}%
%BeginExpansion
\operatorname{eV}%
%EndExpansion
$). Acquisition of this curve is therefore quite gentle. The measured adhesion
force allows to estimate the product of tip radius and contact angle according
to\cite{Israelachvili,AdhesionOthmar}%
\[
F_{ad}=4\pi\gamma R\cos\left(  \varphi\right)
\]
where $R$ is the tip radius, $\gamma$ the surface energy of water and
$\varphi$ the effective contact angle of water on the tip-sample system. With
a force constant $c=2%
%TCIMACRO{\unit{N}}%
%BeginExpansion
\operatorname{N}%
%EndExpansion
$/$%
%TCIMACRO{\unit{m}}%
%BeginExpansion
\operatorname{m}%
%EndExpansion
$ and assuming a wetting sample\cite{CommentWetting}, that is, a contact angle
$\varphi\simeq0-30%
%TCIMACRO{\U{b0}}%
%BeginExpansion
{{}^\circ}%
%EndExpansion
$, we obtain an estimated tip radius of 10-15$%
%TCIMACRO{\unit{nm}}%
%BeginExpansion
\operatorname{nm}%
%EndExpansion
$, in good agreement with results obtained from the non-contact measurements
to be discussed below.

The force curve together with the frequency and amplitude distance behavior
allows to separate four regimes that can be present during tip-sample
approximation (see also Fig. 3): first, a true non-contact regime where the
cantilever essentially oscillates with its free resonance frequency, second, a
\textquotedblleft tapping non-contact\textquotedblright\ regime where the
oscillation amplitude decreases even though no mechanical contact between tip
and sample is formed, third, an intermittent contact regime where the tip
still oscillates and dynamically touches the surface, and forth a continuous
contact regime where the cantilever no longer oscillates. In our experiments
we use rather soft cantilevers at small oscillation amplitudes, therefore we
do not observe the intermittent contact regime (see figure 3), in this
case\ the oscillation stops as soon as the tip mechanically touches the
surface for the first time since the adhesion force is larger than the
restoring force of the cantilever\cite{LangmuirGotas,GiessiblParameters}. With
very small oscillation amplitudes also the second \textquotedblleft tapping
non-contact\textquotedblright\ regime may not be observed, then the tip
directly snaps on to the surface before any significant dissipation is detected.

In the non-contact regimes the \textit{interaction images} are processed as
discussed in the previous section to obtain, for each \textit{interaction
image}, three curves describing the Van der Waals interaction, the capacitance
and the surface potential. Figure 4 shows the result of this process for the
\textit{interaction images} shown in Fig. 2. The graphs in Fig. 4a, 4c and 4e
represent the curves obtained from the force \textit{interaction image} while
the graphs in Fig. 4b, 4d and 4f have been obtained from the frequency
\textit{interaction image}. The Van der Waals curve obtained from the force
\textit{interaction image} essentially reproduces the \textit{FvsD} curve
obtained by simple averaging, a small difference is due to the effect of the
electrostatic interaction: while the curve in Fig. 3 is a mixed
electrostatic/Van der Waals curve\cite{CommentMeanForce}, the curve in Fig. 4a
is a \textquotedblleft true\textquotedblright\ Van der Waals \textit{FvsD}
curve. The parameters obtained from the adjustment of the experimental data to
the relations \ref{freqVdW}-\ref{CapacityFrequency} are summarized in Table 1.

To investigate the precision of the method described in the present work two
test have been made. First, the data describing the interaction has been
adjusted to the relations \ref{freqVdW} and \ref{CapacityFrequency} allowing
the fit to find the exponents $n_{vdW}$ (which was $n_{vdW}=$3 in relation
\ref{freqVdW}) and $n_{el}$ (which was $n_{el}=$2 in relation
\ref{CapacityFrequency}, in the present context only the first term with the
pole $1/(d-d_{0}^{est2})^{2}$ is considered relevant). The results of the
corresponding fits are listed in table 2, confirming that the data indeed
reproduces correctly the exponent of Van der Waals as well as electrostatic interaction.

For the second test the data corresponding to tip-sample capacity (that is,
$C^{\prime\prime}(d)$) was adjusted using two different functions in order to
check whether the data is able to \textquotedblleft see\textquotedblright\ the
difference. One relation is obtained from the usual spherical approximation
which gives $C^{\prime\prime}\left(  d\right)  =2\pi~\epsilon_{0}~R/d^{2}$,
and the second relation is obtained using the more sophisticated model leading
to equation \ref{CapacityFrequency}. In this second model, the cone angle was
fixed to $\vartheta_{0}=20%
%TCIMACRO{\U{b0}}%
%BeginExpansion
{{}^\circ}%
%EndExpansion
$, as specified by the manufacturer. Both models therefore have the tip-radius
as only free parameter. The corresponding fits together with their errors are
shown in Fig. 4d. Even though both fits yield parameters that are compatible
one with the other, the first model for the tip-sample capacitance shows a
clear tendency of the error not observed in the second one.

Figure 5 shows the result of the processing of force and frequency
\textit{interaction images} as well as an oscillation amplitude image acquired
on an Au(111) surface with a Pt-coated Silicon tip. As in the previous case,
data was acquired in ambient air. The graph in Fig. 5a, and 5c represent the
curves obtained from the force \textit{interaction image} while the graph in
Fig. 5b and 5d have been obtained from the frequency \textit{interaction
image}. Table 3 summarizes the values for tip radius and surface position
obtained from the adjustment of the curves 5b-5d to the relations
\ref{freqVdW}-\ref{CapacityFrequency}. Due to the lower oscillation amplitude
(1nm), no "tapping non-contact" range is observed. The mean jump distance is
again about 3.5 nm, and the effective jump distance is 2.5 nm, somewhat larger
than in the previous case. The results obtained from this second experiment
are similar to those obtained in the first one, as can be verified by
comparing the parameters for tip radius and surface position listed in Table 1
and 3. The slightly higher noise level in the second experiment as compared to
the first one is attributed to the lower oscillation amplitude. Nevertheless,
at present we believe that for low oscillation amplitudes our measurements are
limited by technical noise rather than fundamental limits such as shot noise
or thermal movement of the cantilever. In the case of the Si-tip, the Van der
Waals radius ($R^{vdW}$ =5.8$\pm$0.4$%
%TCIMACRO{\unit{nm}}%
%BeginExpansion
\operatorname{nm}%
%EndExpansion
$) is considerably smaller than the two electrostatic radii ($R^{estat1}%
$=12$\pm$1$%
%TCIMACRO{\unit{nm}}%
%BeginExpansion
\operatorname{nm}%
%EndExpansion
$, $R^{estat2}$=21$\pm$4$%
%TCIMACRO{\unit{nm}}%
%BeginExpansion
\operatorname{nm}%
%EndExpansion
$). Moreover, these two electrostatic radii do not agree within the measured
experimental error. One possible explanation for this may be that, at the
level of precision of the method, the Si-tip is not a sufficiently ideal tip
due to the presence of oxide layers on its surface and/or to band bending
effects. In the case of the Pt-tip, the Van der Waals radius ($R^{vdW}$
=14$\pm$5$%
%TCIMACRO{\unit{nm}}%
%BeginExpansion
\operatorname{nm}%
%EndExpansion
$) as well as both electrostatic radii ($R^{estat1}$=16$\pm$12$%
%TCIMACRO{\unit{nm}}%
%BeginExpansion
\operatorname{nm}%
%EndExpansion
$, $R^{estat2}$=21$\pm$6$%
%TCIMACRO{\unit{nm}}%
%BeginExpansion
\operatorname{nm}%
%EndExpansion
$) are compatible one with another, indicating that Van der Waals and
electrostatic interaction "see" the same geometry of the tip. Finally, we note
that the Van der Waals radii obtained for the Si-tip and for the Pt-tip are in
good agreement with the nominal values of the manufacturer: the Si-tip has a
nominal radius better than 10$%
%TCIMACRO{\unit{nm}}%
%BeginExpansion
\operatorname{nm}%
%EndExpansion
$, while the Pt-tip, consisting of a thin 10-15$%
%TCIMACRO{\unit{nm}}%
%BeginExpansion
\operatorname{nm}%
%EndExpansion
$ thin Pt-film on the same kind of Si-tip should have a tip radius of 15-25$%
%TCIMACRO{\unit{nm}}%
%BeginExpansion
\operatorname{nm}%
%EndExpansion
$\cite{Cantilever}.

As a general result from both experiments we find that the data obtained from
the frequency image has less dispersion and less error in the estimation of
the parameters describing the tip-sample interaction. The noise in the data
characterizing the surface potential decreases for small tip-sample distances
only for the data obtained from the frequency \textit{interaction imag}e
(compare figures 4f and 4e). The offset $b_{1}$ obtained from the force
interaction image is clearly different from zero, in fact, it accounts for
almost 50\% of the electrostatic interaction at the smallest tip-sample
distance, while the corresponding offset $b_{2}$ obtained from the frequency
\textit{interaction image} is essentially compatible with $b_{2}=0$. These
offsets reflect the amount of electrostatic interaction that does not change
on a length scale comparable to the distance sampled by an \textit{interaction
image} (100$%
%TCIMACRO{\unit{nm}}%
%BeginExpansion
\operatorname{nm}%
%EndExpansion
$). Therefore, as discussed previously, we conclude that the force signal with
the large value of the offset $b_{1}$ has a significant long range component
of electrostatic interaction while the frequency, with its vanishing offset
$b_{2}$, has a truly short range behavior. As described in detail
elsewhere\cite{modeloPRB}, the long range component of the electrostatic
interaction is due to the mesoscopic tip cone or the macroscopic cantilever,
while the short range interaction is induced by the nanoscopic tip apex,
usually described by means of an (effective) tip radius. In the frequency
(that is, force gradient) signal, the long range component of the interaction
is "derived away" and only the interaction with the nanoscopic tip apex is
relevant. In our opinion, this explains the large value of $b_{1}$ obtained
from the force \textit{interaction image} as compared to the vanishing of
$b_{2}$ obtained from the frequency \textit{interaction image}.

For small indentation forces, the experiments described here are quite
reproducible. This is seen in Fig. 6, which shows the result of processing
three different \textit{interaction images} taken consecutively at a same spot
of an Au(111) surface. The corresponding \textit{FvsD} curves obtained from
force \textit{interaction images} are shown with approach and retraction cycle
(Fig. 6a). The second derivative of the capacitance as well as the contact
potential were obtained from frequency \textit{interaction images} and only
approach curves are presented. Within the experimental error the three
indentations lead to essentially the same results. In particular, the
\textit{FvsD} curves are almost absolutely equivalent, the three curves show
not only the same jump to contact distance and the same adhesion force, also
the details of how the tip-sample contact breaks are very similar. Images
taken before and after the acquisition of these curves showed no variation of
the surface, confirming that the surface was not modified during the
indentation experiments. When thermal drift is sufficiently low, an even more
gentle measurement of tip-sample interaction is possible by acquiring
\textit{interaction images} without bringing the tip into mechanical contact
with the sample. Tip-sample distance can then in principle be estimated from
the values of the poles $d_{0}$ of tip-sample interaction (see relations
\ref{freqVdW}-\ref{CapacityFrequency}). In this case, the tip is maintained
pristine. In particular when very sharp tips are used, this is a significant
advantage over other techniques for measuring tip-sample interaction. In our
opinion, the results shown in Fig. 6 demonstrate the potential and precision
of the method. The method is sufficiently sensitive to detect possible
variations of the tip-sample system, which may be detected in any of the
measured parameters, that is, not only in the \textit{FvsD} curve (and in
particular in the adhesion force), but also in the surface potential or in the
capacitance of the tip-sample system.

\section{Conclusion}

In the present work, the Van der Waals and electrostatic contribution to the
interaction in a SFM set-up have been separated and analyzed quantitatively.
The method described is based on the acquisition of \textit{interaction
images}\ where the force, the oscillation amplitude and the resonance
frequency are acquired as a function of tip-sample distance and tip-sample
voltage. Using appropriate processing algorithms, the Van der Waals
interaction, the electrostatic interaction and the contact potential are
determined from the force as well as from the frequency data. We note in this
context that other methods for measuring tip-sample interaction do not
directly allow the separation of Van der Waals and electrostatic interaction.
In particular, when the materials of tip and sample are different -which is
usually the case- contact potentials may significantly contribute to the total
tip-sample interaction. If pure Van der Waals interaction is to be measured,
the contact potential has to be compensated and if electrostatic interaction
is measured, the contact potential has to be taken into account. With the
method described here, Van der Waals interaction, electrostatic interaction -
described by means of a tip-sample capacitance or some of its derivatives -
and contact potentials are determined \textquotedblleft self
consistently\textquotedblright, that is, the method automatically takes
account of the effect induced by each of these three phenomena. More
technically, our method assumes no prior information about the physical
properties of the tip-sample system, instead it "finds" the combination of
parameters corresponding to Van der Waals and electrostatic interaction that
best describe the measured data.

Our experiments performed with (highly doped) semi-conducting as well as with
metalized tips demonstrate the potential of the method and yield high
precision, quantitative and reproducible interaction data. The high precision
of the method is due to the fitting process which implies averaging of
experimental data. Typically 100 to 1000 data points are acquired (one line of
an \textit{interaction image}) to obtain three physically relevant parameters,
therefore the signal to noise ratio of these curves increases as compared to
the usual acquisition of interaction versus distance curves. Assuming that
data fitting increases the signal to noise ratio as $\sqrt{n}$, with $n$
number of data points, the improvement of signal to noise ratio is about one
order of magnitude. Another feature of the method presented is that the
processing algorithms can be developed further to test whether the
electrostatic interaction follows the quadratic dependence with bias voltage
described by relation \ref{SFMforce}. This is particularly relevant for
semiconducting tips or samples as well as for strong electric fields between
tip and sample where departures from the quadratic behavior could be expected.
In addition, also inhomogeneities of the tip surface potential could be
recognized due to non-ideal behavior of the interaction with tip-sample
voltage as well as due to a distance dependence of the measured surface
potential (see also\cite{modeloPRB}).

We have shown that the technique is able to experimentally determine the
exponent of the Van der Waals and the electrostatic interaction. In addition,
the technique is also able to differentiate between two models for
electrostatic tip-sample interaction, showing that an accurate description of
electrostatic interaction needs to go beyond the simple sphere-sample
approximation used generally.

We believe that the method presented in this work will significantly
contribute to improve the measurement of interaction in a SFM setup and aid
the correct interpretation of SFM experiments. In addition, an improved
characterization of interaction will allow quantitative determination of
material properties on a nanometer scale, an issue that recently is receiving
more and more attention but that, in our opinion, still needs some important
efforts from the SPM community.

\section{Acknowledgments}

The authors acknowledge stimulating discussions with J. Abellan, A. Urbina, C.
Munuera, J. G\'{o}mez, C. G\'{o}mez-Navarro, A.M. Bar\'{o} and A. Gil. The
authors also thank Atomic Force F\&E GmbH, and in particular Mr. Ludger
Weisser,\ for supplying the cantilevers\ used. M. Cuenca, I. Horcas, P.
Colilla and R. Fernandez from the Nanotec team contributed to this work by
adapting instrumental design and software routines. And finally, the work was
supported by the Spanish Ministry of Science and Technology as well as by the
Fundaci\'{o}n S\'{e}neca-CARM through the projects MAT2002-01084,
NAN-2004-09183C10-03 and "Crecimiento de Nanoestructuras....".

\section{Tables}

\begin{tabular}
[c]{||l|l|l|l||}\hline\hline
curve & tip radius $R$ & surface position $d_{0}$ & offset $b$\\\hline
force $F(d)$ & $10-15nm$ (from $F_{ad}$) & $0$ (by definition) & $-$\\\hline
frequency $\nu(d)$ & $R^{vdW}=5.8\pm0.4nm$ & $d_{0}^{VdW}=3.1\pm0.2nm$ &
$-$\\\hline
force-capacitance $C^{\prime}(d)$ & $R^{estat1}=12\pm4nm$ & $d_{0}%
^{estat1}=2\pm2nm$ & $b_{1}=-(7\pm4)\ast10^{-3}$\\\hline
frequency-capacitance $C^{\prime\prime}(d)$ & $R^{estat2}=21\pm1nm$ &
$d_{0}^{estat2}=1.2\pm0.9nm$ & $b_{2}=-(37\pm14)\ast10^{-6}$\\\hline\hline
\end{tabular}

Table 1

Parameters describing tip-sample interaction obtained after processing the
\textit{interaction images} shown in Fig. 2, and the curves shown in Figs.
4b), 4c) and 4d). Data acquired with a Si-tip on an Au (111) surface.
Resonance frequency: 73.67 kHz, force constant 2N/m, free oscillation
amplitude $2.5%
%TCIMACRO{\unit{nm}}%
%BeginExpansion
\operatorname{nm}%
%EndExpansion
$.\bigskip

\begin{tabular}
[c]{||l|l|l|l|l||}\hline\hline
Curve & tip radius $R$ & surface position $d_{0}$ & exponent experiment &
exponent theory\\\hline
$\nu(d)$ & $15\pm30nm$ & $3\pm2nm$ & $3.3\pm0.5$ & $3$\\\hline
$C^{\prime\prime}(d)$ & $17\pm1nm$ & $2.2\pm0.2nm$ & $1.9\pm0.5$ &
$2$\\\hline\hline
\end{tabular}
\bigskip

Table 2

Parameters describing tip-sample interaction obtained after processing the
\textit{interaction images} shown in Fig. 2, and the curves shown in Figs.
4b), 4c) and 4d). In addition to the parameters $R$ and $d_{0}$, also the
exponents describing the variation with tip-sample distance in relations
\ref{CapacityFrequency} and \ref{freqVdW} were allowed to vary.\bigskip

\begin{tabular}
[c]{||l|l|l|l||}\hline\hline
curve & tip radius $R$ & surface position $d_{0}$ & offset $b$\\\hline
force $F(d)$ & $10-15nm$ (from $F_{ad}$) & $0$ (by definition) & $-$\\\hline
frequency $\nu(d)$ & $R^{vdW}=14\pm5nm$ & $d_{0}^{VdW}=1.7\pm0.5nm$ &
$-$\\\hline
force-capacitance $C^{\prime}(d))$ & $R^{estat1}=16\pm12nm$ & $d_{0}%
^{estat1}=2.1\pm1.1nm$ & $b_{1}=-0.1\pm0.3$\\\hline
frequency-capacitance $C^{\prime\prime}(d)$ & $R^{estat2}=21\pm6nm$ &
$d_{0}^{estat2}=1.2\pm0.7nm$ & $b_{2}=-(1.7\pm1.3)\ast10^{-4}$\\\hline\hline
\end{tabular}
\bigskip

Table 3

Parameters describing tip-sample interaction obtained after processing
\textit{interaction images} acquired with a Pt-tip on an Au (111) surface.
Resonance frequency: 68.80 kHz, force constant 2N/m, oscillation amplitude $1%
%TCIMACRO{\unit{nm}}%
%BeginExpansion
\operatorname{nm}%
%EndExpansion
$.\bigskip

\bigskip

\section{Figure Captions}

\begin{center}
\bigskip%
%TCIMACRO{\FRAME{itbpF}{16.1671cm}{6.6689cm}{0cm}{}{}{fig1_new.jpg}%
%width 16.1671cm;  height 6.6689cm;  depth 0cm;  original-width 13.7194in;
%original-height 6.4965in;  cropleft "-0.0774";  croptop "1";
%cropright "1.0774";  cropbottom "0";
%filename 'Figs/Fig1_New.jpg';file-properties "XNPEU";}}}%
%BeginExpansion
\raisebox{-0cm}{\includegraphics[
trim=-1.061882in 0.000000in -1.061881in 0.000000in,
natheight=6.496500in,
natwidth=13.719400in,
height=6.6689cm,
width=16.1671cm
]%
{Figs/Fig1_New.jpg}%
}%
%EndExpansion

\end{center}

Figure 1

Schematic description of the experimental setup used to acquire
\textquotedblleft interaction images\textquotedblright. The lateral position
of the tip over the sample is fixed. The \textquotedblleft
fast\textquotedblright\ and \textquotedblleft slow\textquotedblright\ outputs
of the electronics are used to vary the tip-sample voltage and the tip-sample
distance. A Phase Locked Loop is used to mechanically oscillate the tip-sample
system always at resonance. Frequency shifts $\Delta v$ are measured with
respect to a reference frequency -usually the free resonance frequency- set by
the electronic system. In our experiments, the force, the frequency shifts and
the variation of oscillation amplitude induced by the tip-sample interaction
are measured simultaneously.

\begin{center}
\bigskip%
%TCIMACRO{\FRAME{itbpF}{1.8308in}{5.8323in}{0in}{}{}{fig2_new.jpg}%
%width 1.8308in;  height 5.8323in;  depth 0in;  original-width 4.5065in;
%original-height 14.5133in;  cropleft "0";  croptop "1";  cropright "1";
%cropbottom "0";  filename 'Figs/Fig2_New.jpg';file-properties "XNPEU";}}}%
%BeginExpansion
{\includegraphics[
natheight=14.513300in,
natwidth=4.506500in,
height=5.8323in,
width=1.8308in
]%
{Figs/Fig2_New.jpg}%
}%
%EndExpansion

\end{center}

Figure 2

Typical set of interaction images acquired by varying the bias voltage
(\textquotedblleft fast\textquotedblright\ scan, corresponding to a horizontal
scan line) and the tip-sample distance (\textquotedblleft
slow\textquotedblright\ scan, vertical direction). The upper part of the
images corresponds to large tip-sample distances, the lower part to small
distances. At the bottom of the images, tip and sample are in mechanical
contact. Resonance frequency: 73.67 kHz, force constant 2N/m, oscillation
amplitude $2.5%
%TCIMACRO{\unit{nm}}%
%BeginExpansion
\operatorname{nm}%
%EndExpansion
$.

a) Force \textquotedblleft interaction image\textquotedblright, total gray
scale corresponds to about 1 nN.

b) Frequency \textquotedblleft interaction image\textquotedblright, total gray
scale corresponds to about 300 Hz.

c) Oscillation amplitude \textquotedblleft interaction image\textquotedblright%
, total gray scale corresponds to a variation of 0.5~nm (that is about 20\% of
the total oscillation amplitude). All images have been taken simultaneously
with a voltage scan of $\pm$1.5V, and the total tip movement is 100nm.
Clearly, the parabolic dependence of interaction on the bias voltage is
recognized best in the frequency image for small tip-sample distances.

\begin{center}
\bigskip%
%TCIMACRO{\FRAME{itbpF}{7.7321cm}{4.67cm}{0cm}{}{}{fig3_new.jpg}%
%width 7.7321cm;  height 4.67cm;  depth 0cm;  original-width 13.3864in;
%original-height 9.0529in;  cropleft "-0.0633";  croptop "1";
%cropright "1.0633";  cropbottom "0";
%filename 'Figs/Fig3_New.jpg';file-properties "XNPEU";}}}%
%BeginExpansion
\raisebox{-0cm}{\includegraphics[
trim=-0.847359in 0.000000in -0.847359in 0.000000in,
natheight=9.052900in,
natwidth=13.386400in,
height=4.67cm,
width=7.7321cm
]%
{Figs/Fig3_New.jpg}%
}%
%EndExpansion

\end{center}

Figure 3

Force vs. distance curve calculated from the force and amplitude
\textit{interaction images}\ shown in Fig. 2. From the force
\textit{interaction image} the mean deflection measured by the detection
system during the corresponding line of the \textit{interaction image}\ is
shown as solid circles (approach) and triangles (retraction). From the
amplitude \textit{interaction image}\ the oscillation amplitude is calculated
and represented as the upper and lower lines in the graph showing the upper
and lower turning point of the cantilever during the oscillation. As discussed
in the main text, the piezo displacement has been adjusted in order to have
the position $\Delta=0$ at the point where the deflection of the cantilever is
zero when tip and sample are in contact. The (apparent) jump distance is about
$4%
%TCIMACRO{\unit{nm}}%
%BeginExpansion
\operatorname{nm}%
%EndExpansion
$ and the adhesion force is about 9 nN (=4.5nm x 2N/m). The different
interaction regimes described in the main text are separated with dashed
lines: (I) "true non-contact regime", (II) "tapping non-contact" and (IV)
"continuous contact regime". Due to the soft cantilevers and the low
oscillation amplitude used, the "intermittent contact regime" (referred to as
region III in the main text) does not appear.

\begin{center}
\bigskip%
%TCIMACRO{\FRAME{itbpF}{12.6855cm}{11.4664cm}{0cm}{}{}{fig4_new.jpg}%
%width 12.6855cm;  height 11.4664cm;  depth 0cm;  original-width 15.0469in;
%original-height 13.594in;  cropleft "0";  croptop "1";  cropright "1";
%cropbottom "0";  filename 'Figs/Fig4_New.jpg';file-properties "XNPEU";}}}%
%BeginExpansion
\raisebox{-0cm}{\includegraphics[
natheight=13.594000in,
natwidth=15.046900in,
height=11.4664cm,
width=12.6855cm
]%
{Figs/Fig4_New.jpg}%
}%
%EndExpansion

\end{center}

Figure 4

Several interaction\ vs. distance curves calculated from the force and
frequency \textit{interaction images}\ as discussed in detail in the main
text. Graphs b), c) and d) show not only the data points, but also the best
fit to the data points using the relations \ref{freqVdW}%
-\ref{CapacityFrequency} as well as the error to the best fit (smaller points
around the dashed horizontal line). For the frequency data, this error is very
small, therefore an amplification factor of 10 has been applied as compared to
the data points.

Left graphs: Curves calculated from the force \textit{interaction image}: a)
Van der Waals force vs. distance curve, c) first derivative of the tip-sample
capacitance vs. distance together with the best fit to the data and the
corresponding error and e) surface potential vs. distance.

Right graphs: Curves calculated from the frequency \textit{interaction image}:
b) Van der Waals frequency vs. distance curve together with the best fit to
the data and the corresponding error (x10), d) second derivative of the
tip-sample capacitance vs. distance together with the best fit to the data and
the corresponding error (x10), and f) surface potential vs. distance. For the
capacity data calculated from the frequency \textit{interaction image} (graph
d)) the error is shown for two different relations, corresponding to two
different descriptions of the tip-sample system. One of the two models (the
empty circles) shows a small but clear tendency of the error.

In all graphs, the piezo displacement has been adjusted with the value used in
Fig. 3. Note that only data points corresponding to the non-contact regime are
shown, therefore no data is shown for tip-sample distances $\Delta<5$nm.
Resonance frequency: 73.67 kHz, force constant 2N/m, oscillation amplitude
$2.5%
%TCIMACRO{\unit{nm}}%
%BeginExpansion
\operatorname{nm}%
%EndExpansion
$.

\begin{center}
\bigskip%
%TCIMACRO{\FRAME{itbpF}{16.6153cm}{8.7426cm}{0cm}{}{}{fig5_new.jpg}%
%width 16.6153cm;  height 8.7426cm;  depth 0cm;  original-width 15.3132in;
%original-height 10.3466in;  cropleft "-0.1445";  croptop "1";
%cropright "1.1445";  cropbottom "0";
%filename 'Figs/Fig5_New.jpg';file-properties "XNPEU";}}}%
%BeginExpansion
\raisebox{-0cm}{\includegraphics[
trim=-2.212757in 0.000000in -2.212758in 0.000000in,
natheight=10.346600in,
natwidth=15.313200in,
height=8.7426cm,
width=16.6153cm
]%
{Figs/Fig5_New.jpg}%
}%
%EndExpansion

\end{center}

Figure 5

Interaction vs. distance curves calculated from force and frequency
\textit{interaction images}\ (not shown) acquired with a Pt-tip on an Au(111)
surface in air. As in Figure 4, graphs b), c) and d) show not only the data
points, but also the best fit to these data points using the relations
\ref{freqVdW}-\ref{CapacityFrequency}. The error to the best fit is
represented as smaller points around the dashed horizontal line. In this case,
no amplification factor has been applied to the errors.

Left graphs: Curves calculated from the force \textit{interaction image}: a)
(mean) \textit{FvsD} curve, as in Figure 3 forward and backward cycle are
shown. The thicker points correspond to the calculated mean force while the
two lines below and over these data points show the oscillation amplitude
calculated from the amplitude \textit{interaction image}. Note that, in
contrast to figure 3, only the "true non-contact regime" (region I in figure
3)\ and the continuous contact regime (region IV in figure 3) are measured due
to the small oscillation amplitude. c) first derivative of the tip-sample
capacitance vs. distance.

Right graphs: Curves calculated from the frequency \textit{interaction image}:
b) (Van der Waals) frequency vs. distance curve and d) second derivative of
the tip-sample capacitance vs. distance. Resonance frequency: 68.80 kHz, force
constant 2N/m, oscillation amplitude 1nm.

\begin{center}
\bigskip%
%TCIMACRO{\FRAME{itbpF}{7.8419cm}{11.1808cm}{0cm}{}{}{fig6_new.jpg}%
%width 7.8419cm;  height 11.1808cm;  depth 0cm;  original-width 10.1996in;
%original-height 14.5799in;  cropleft "0";  croptop "1";  cropright "1";
%cropbottom "0";  filename 'Figs/Fig6_New.jpg';file-properties "XNPEU";}}}%
%BeginExpansion
\raisebox{-0cm}{\includegraphics[
natheight=14.579900in,
natwidth=10.199600in,
height=11.1808cm,
width=7.8419cm
]%
{Figs/Fig6_New.jpg}%
}%
%EndExpansion

\end{center}

Figure 6

Interaction vs. distance data calculated from three consecutive
\textquotedblleft interaction images\textquotedblright\ acquired with a Pt-tip
on a Au(111) surface. The data obtained from the three interaction images;
the\ first indentation correspond to circles, the second to diamonds and the
third one to triangles. Top image (a): (mean) \textit{FvsD} curve: data
corresponding to tip-sample approach and tip-sample retraction is shown. Note
that the data corresponding to tip-sample retraction, that is, the part
showing adhesion, is represented without filling to aid the eye. Central image
(b): Second derivative of tip-sample capacitance obtained from the frequency
\textit{interaction images}. Only approach data is shown, the solid circles
correspond to the first indentation, the gray diamonds to the second and the
white triangles to the third one. Lower image (c): Surface potential
calculated from the frequency \textit{interaction images}. Again only approach
data is shown and the coding for the different indentations is the same as for
graph (b). Resonance frequency: 68.80 kHz, force constant 2N/m, oscillation
amplitude 1nm.


\begin{thebibliography}{99}
\bibitem {CommentNonContact1}Here \textquotedblleft
non-contact\textquotedblright\ regime refers to the case where tip and sample
are neither in mechanical contact nor in very close proximity to the sample,
that is, for distances larger than about 1nm in UHV and 2-5 nm in ambient air.
For smaller distances chemical forces and -once mechanical contact is
established- also elastic forces have to be considered. In addition, in air
the interaction due to the formation of liquid necks has to be taken into
account. In the present context \textquotedblleft true" non-contact\ regime
will refer to the regime where only Van der Waals and electrostatic forces are relevant.
\bibitem {CommentCapacity}Strictly, only the first and second derivative of
tip-sample capacitance are determined, but not the capacitance itself.
\bibitem {FrequencyApprox}To be more precise: the effective force constant
$c_{eff}=c-\partial F\left(  d\right)  /\partial d$\ of the system should not
vary significantly over the range of cantilever oscillation. This implies that
$\,\left\vert F^{\prime\prime}(d_{0})\right\vert a\ll c_{lever}$, where
$a$\ is the oscillation amplitude, $F^{\prime\prime}(d_{0})$\ the second
derivative of the force at the mean equilibrium position $d_{0}$, and
$c_{lever}$\ is the force constant of the cantilever.
\bibitem {CommentCali}That is, the known piezo calibration is transferred to a
calibration of the photodiode by assuming a slope of 1 for the contact regime
of the cantilever deflection vs. distance curve. For more details, see ref.
\cite{FvsDWeissenhorn}.
\bibitem {CommentLockRange}Since $\nu_{cont}\simeq4.4~\nu_{free}$, with our
free resonance frequency of $\nu_{free}\simeq68kHz$ the resoncance frequency
in contact is not within the lock range of about $15kHz$ of our PLL).
\bibitem {CommentWetting}Prior to the SFM\ experiments, the Au-(111) surface
was flame annealed to remove contaminants. To estimate the contact angle a
small drop of water was deposited on an edge of the sample and found to
spread, indicating that sample is macrosopically wetting and that the amount
of contamination due to hydrocarbons is relatively low.
\bibitem {CommentMeanForce}Force data is averaged over a whole "fast-scan"
line, where the voltage is varied between $\pm U_{scan}$. Therefore in
addition to the Van der Waals interaction, on average also an electrostatic
force is acting, since the electrostatic force is proportional to
$U_{scan}^{2}$, and does not vanish over a"fast-scan" line.
\end{thebibliography}
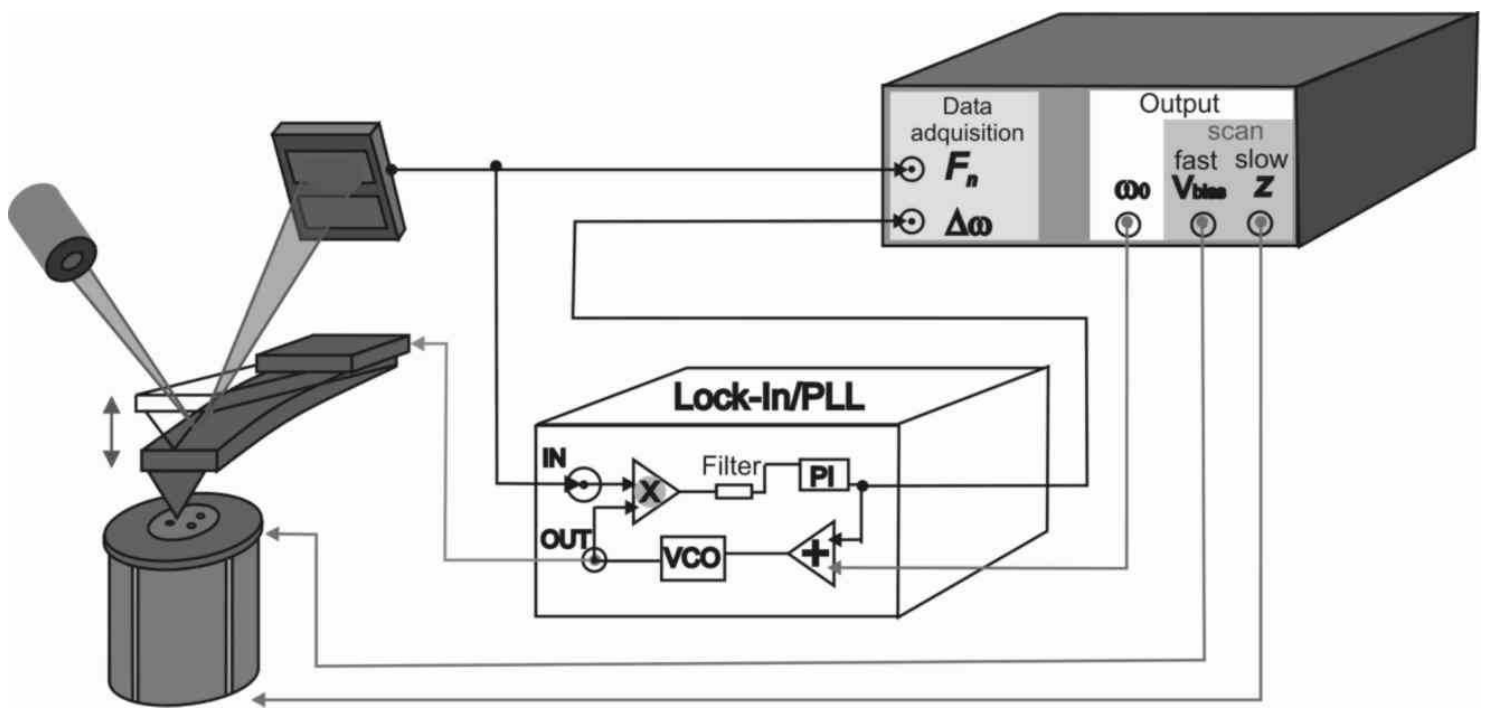
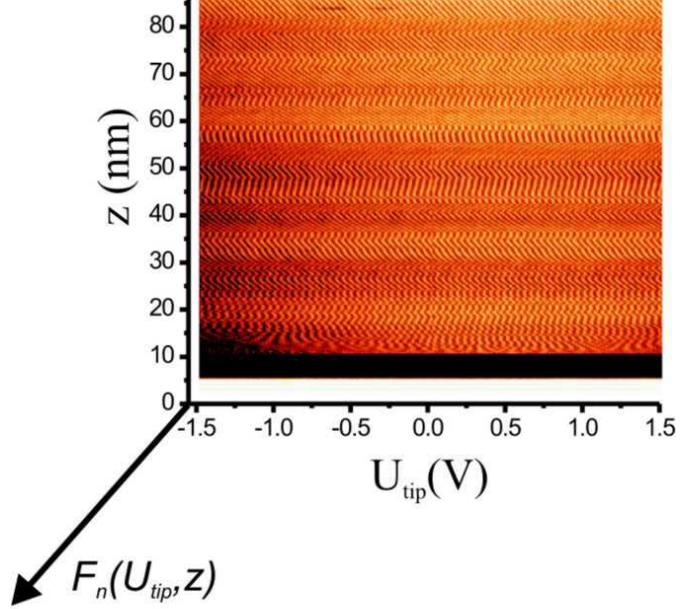
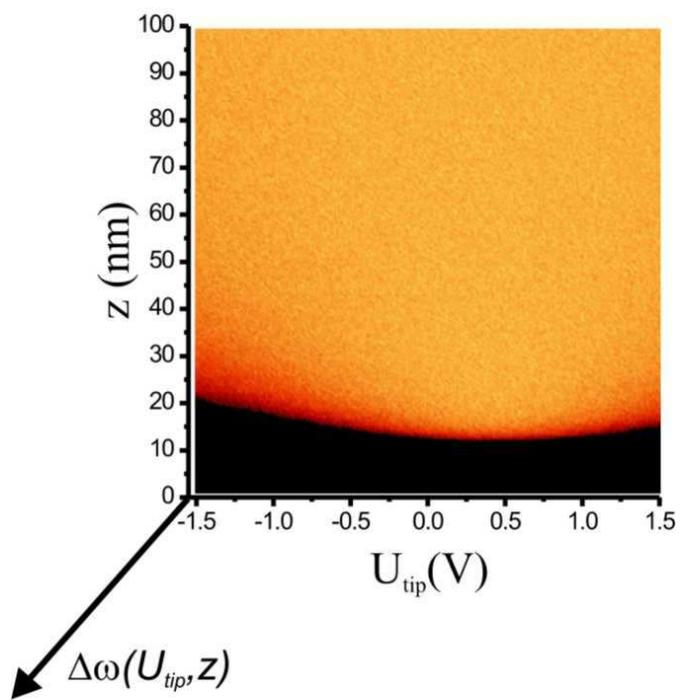
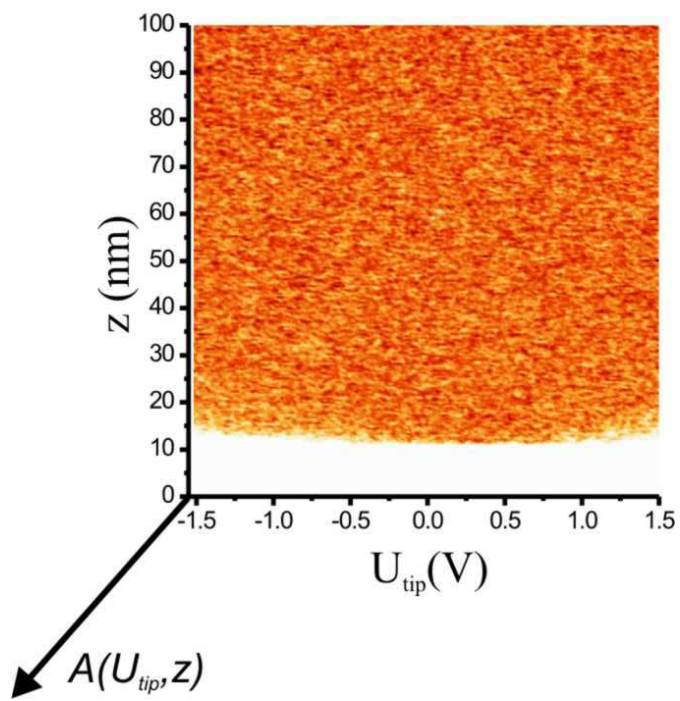
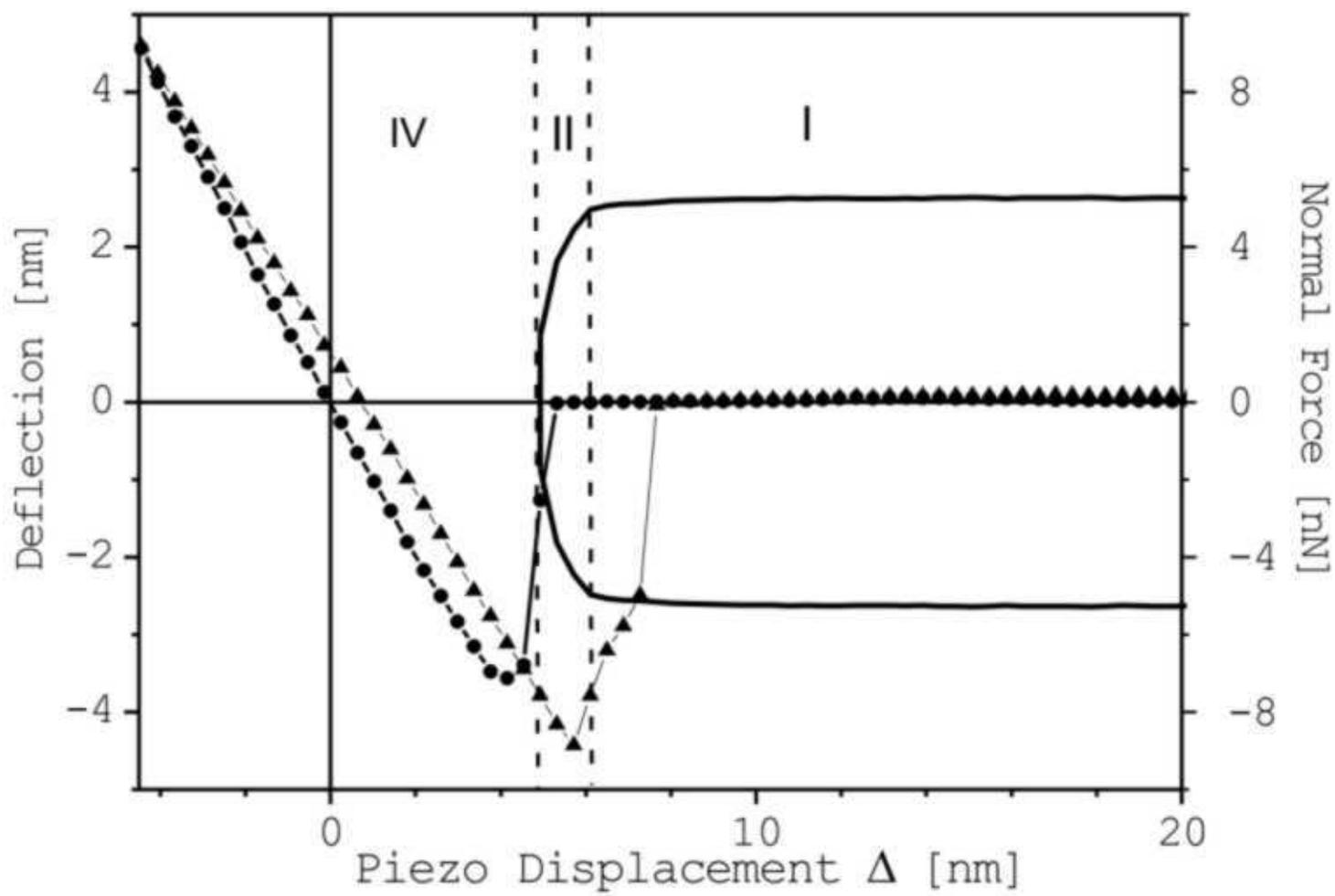
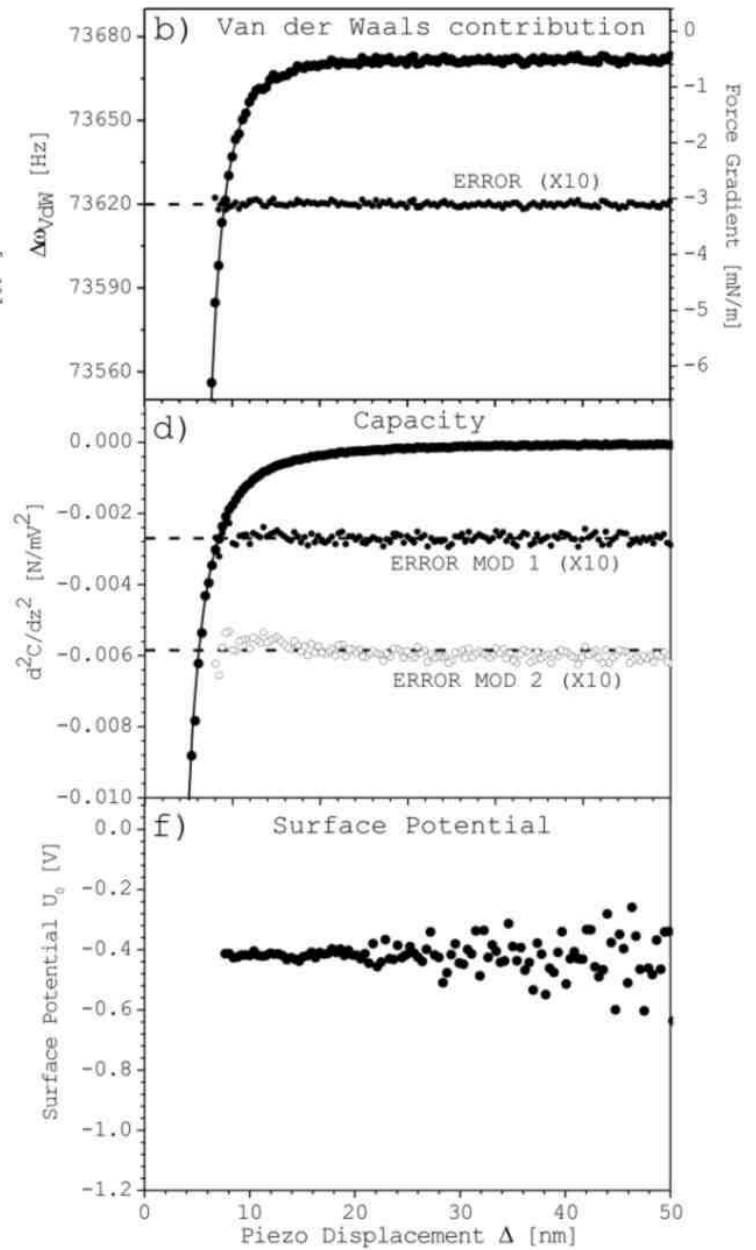
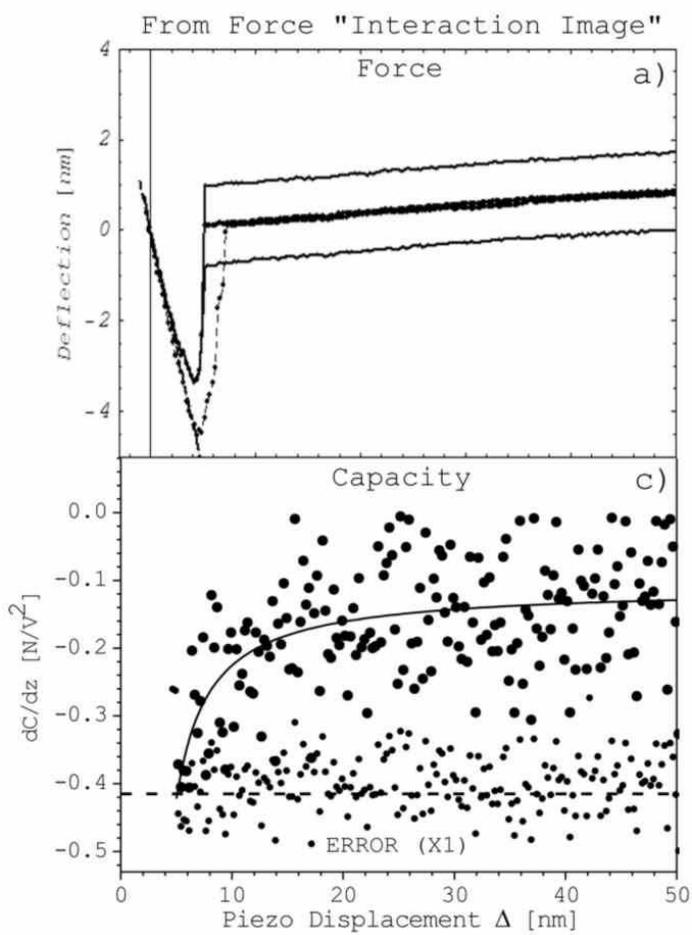
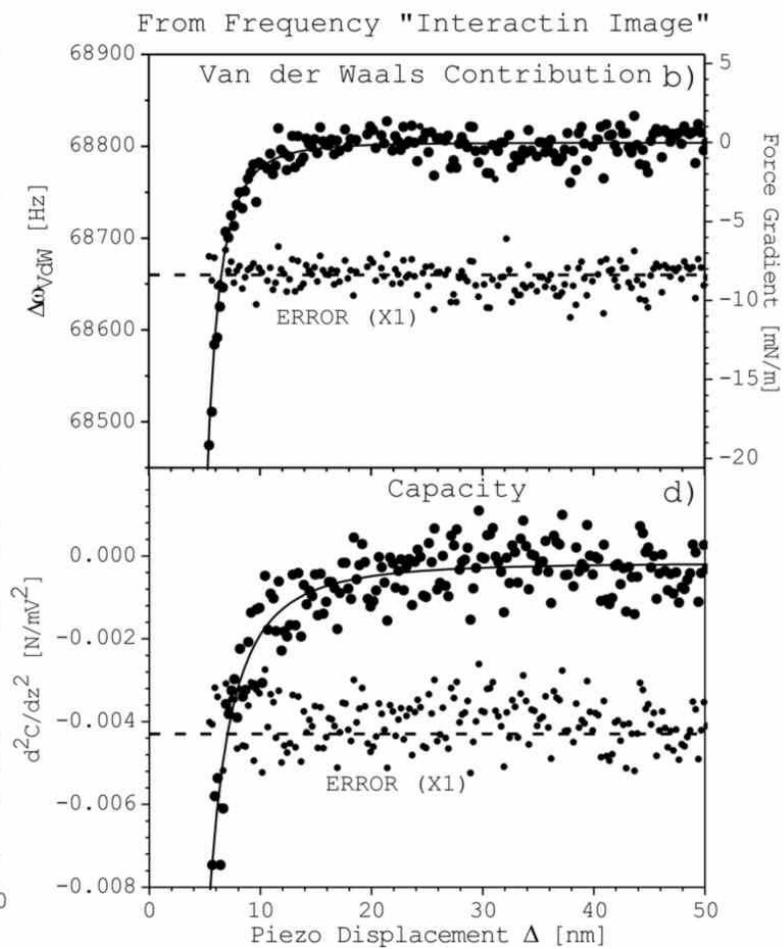
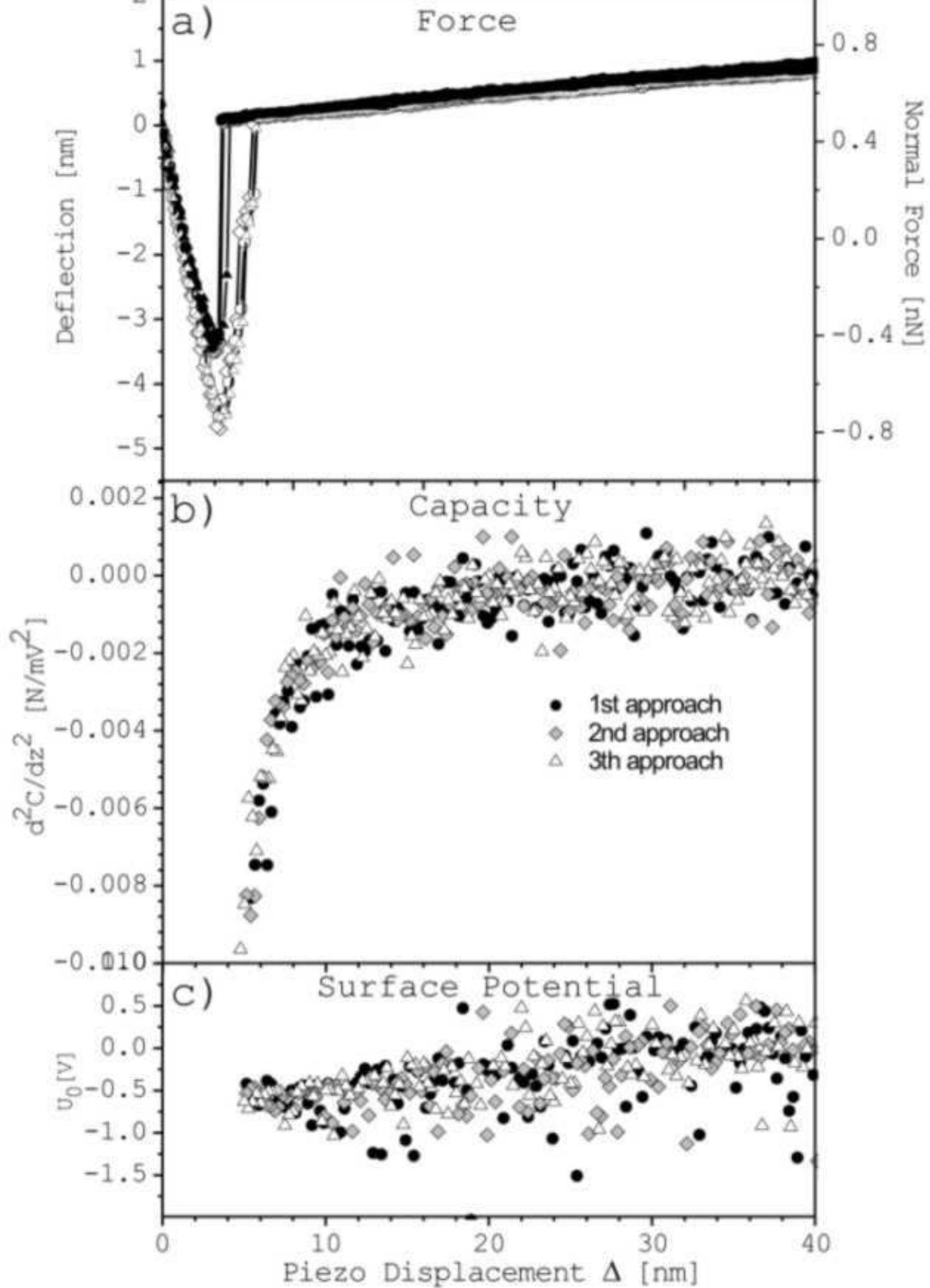
\end{document}